\documentclass [12pt]{article}
\usepackage{graphics,color,epsfig}

\begin{document}

\begin{flushright}
CLEO CONF 01-2\\
\end{flushright}
\bigskip
\begin{center}
\textbf{\Large First Measurement of $\Gamma(D^{\ast +})$}\\

\bigskip

CLEO Collaboration\\

\bigskip

(February 5, 2001)

\end{center}

\begin{abstract}

We have made the first measurement of the $D^{\ast +}$ width using
9/fb of $e^+e^-$ data collected near the $\Upsilon(4\rm{S})$
resonance by the CLEO II.V detector.
Our method uses advanced tracking techniques
and a reconstruction method that takes advantage of the small
vertical size of the CESR beam spot to measure the energy release
distribution from the $D^{\ast +} \to D^0 \pi^+$ decay.
Our preliminary result is $\Gamma(D^{\ast +}) = 96 \pm 4\ ({\rm Statistical})
\pm 22\ ({\rm Systematic})$ keV.

\end{abstract}

\newpage

{
\renewcommand{\thefootnote}{\fnsymbol{footnote}}
\begin{center}
T.~E.~Coan,$^{1}$ V.~Fadeyev,$^{1}$ Y.~S.~Gao,$^{1}$
Y.~Maravin,$^{1}$ I.~Narsky,$^{1}$ R.~Stroynowski,$^{1}$
J.~Ye,$^{1}$ T.~Wlodek,$^{1}$
M.~Artuso,$^{2}$ C.~Boulahouache,$^{2}$
% ...
\end{center}
 
\small
\begin{center}
$^{1}${Southern Methodist University, Dallas, Texas 75275}\\
$^{2}${Syracuse University, Syracuse, New York 13244}\\
% ...
\end{center}

\setcounter{footnote}{0}
}
\newpage

\section{Introduction}

A measurement of $\Gamma(D^{\ast +})$ opens an important window
on the non-perturbative strong physics involving heavy quarks.
The basic framework of the theory is well understood, however, there is
still much speculation -
predictions for the width range from $15\,\rm{keV}$ to 
$150\,\rm{keV}$ \cite{pred}.
We know the $D^{\ast +}$ width is dominated by strong decays since 
the measured electromagnetic transition rate is small,
$(1.68\pm0.45)$\% \cite{mats}.
The level splitting in the $B$ sector
is not large enough to allow real strong transitions.
Therefore, a measurement of the width of the $D^{\ast +}$
gives unique information about the
strong coupling constant in heavy-light systems.

	The total width of the $D^{\ast +}$ is the sum of the partial widths
of the strong decays $D^{\ast +} \to D^0 \pi^+$ and $D^{\ast +} \to D^+ \pi^0$
and the radiative decay $D^{\ast +} \to D^+ \gamma$.  We can write
the width in terms of a single strong coupling, $g$, and an
electromagnetic coupling, $g_{D^+ \gamma}$:
\begin{eqnarray}
\Gamma(D^{\ast +}) & = & 
              \Gamma(D^0 \pi^+) + \Gamma(D^+ \pi^0) + \Gamma(D^+ \gamma) \\
                   & = & \frac{2g^2}{48\pi m_{D^{\ast +}}^2}p^3_{\pi^+} +
                         \frac{ g^2}{48\pi m_{D^{\ast +}}^2}p^3_{\pi^0} +
                         \frac{\alpha g_{D^+ \gamma}^2}{3}p^3_{\gamma}
\end{eqnarray}
using the isospin relationship
\begin{equation}
g = g_{D^+ \pi^0} = g_{D^0 \pi^+}/\sqrt{2}, 
\end{equation}
where $\alpha$ is the fine structure
constant, and the momenta are those for the indicated particle in the
appropriate $D^{\ast +}$ decay in its rest frame.  The contribution
of the electromagnetic decay can be neglected to note that the width
of the $D^{\ast +}$ only depends on $g$ \cite{pred}.

\begin{figure}
\centering
\epsfxsize=14cm
\epsfysize=12cm
\epsfbox{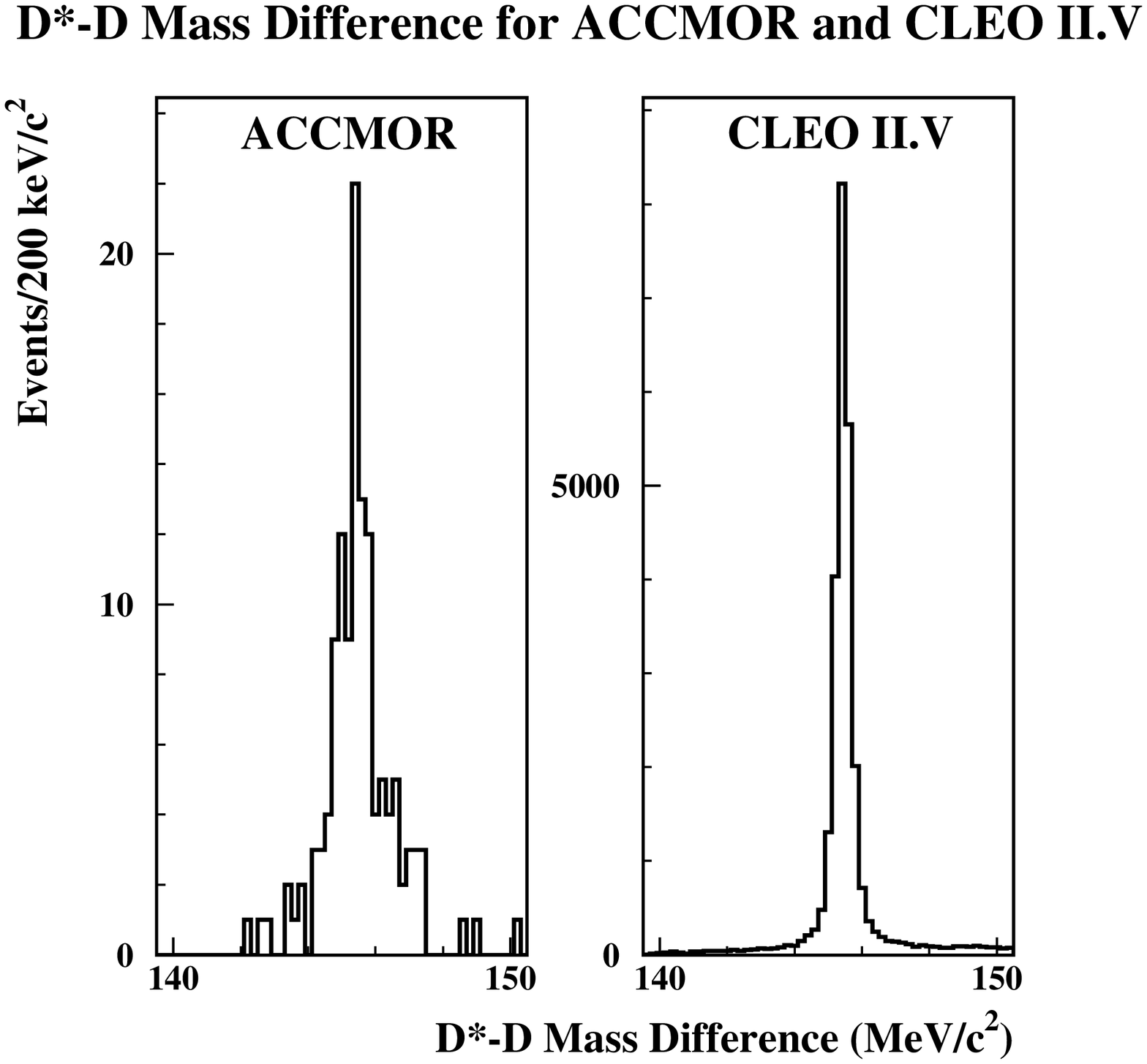}
\caption{\label{fig:widths} The $D^{\ast +}$ - $D^0$ mass difference
distribution obtained by ACCMOR (left){\protect\cite{ACCMOR}} and in this work.}
\end{figure}

Prior to this measurement, the $D^{\ast +}$ width was limited
to be less than $131\,\rm{keV}$ at the $90\%$ confidence level 
by the ACCMOR collaboration \cite{ACCMOR}.
The limit was based on 110 signal events reconstructed in two
$D^0$ decay channels with a background of 15\% of the signal.
This contribution describes
a measurement of the $D^{\ast +}$ width with the CLEO II.V detector
where the signal, in excess of 13,000 events,  is
reconstructed through a single, well-measured sequence,
$D^{\ast +} \rightarrow \pi^+_{\rm slow} D^0$,
$D^0 \rightarrow K^-\pi^+$.  Consideration of
charge conjugated modes are implied throughout this paper.
The level of background under the signal is less than 3\%
in our loosest selection.
Figure~\ref{fig:widths} compares the CLEO II.V signal with signal
from the previously published limit.

The challenge of measuring the width of the $D^{\ast +}$ is understanding the tracking
system response function since the experimental resolution exceeds the width
we are trying to measure. 
Candidates with mismeasured hits, errors in pattern recognition, and
large angle Coloumb scattering are particularly dangerous because
the signal shape they project is broad and the errors for these events can be
underestimated, resulting in events that can easily influence the 
parameters of a Breit-Wigner fitting shape.  We generically term
such effects ``tracking mishaps.''
A difficulty is that there is no physical calibration for this measurement.
The ideal calibration mode would have a large
cross-section, a width of zero, decay with a rather small
energy release to three charged
particles one of which has a much softer momentum distribution than the
other two,
and the other two would decay through a zero width resonance
with a measurable flight distance.
Such a mode would allow us to disentangle detector effects from the underlying
width but no such mode exists.

Therefore, to measure the width of the $D^{\ast +}$ we depend
on exhaustive comparisons between a GEANT \cite{GEANT} based detector
simulation and our data.  We addressed the problem by selecting samples of
candidate $D^{\ast +}$ decays using three strategies.

	First we produced the largest sample from data and simulation by 
imposing only
basic tracking consistency requirements.  We call this the
{\em nominal} sample.

	Second we refine the nominal sample selecting candidates
with the best measured tracks by making very tight cuts
on tracking parameters.  There is special emphasis on choosing those tracks
that are well measured in our silicon vertex detector.  This reduces
our nominal sample by a factor of thirty and, according to our
simulation, has negligible contribution from tracking mishaps.
We call this the {\em tracking selected} sample.

	A third alternative is to select our data on specific kinematic
properties of the $D^{\ast +}$ decay that minimize the dependence
of the width of the $D^{\ast +}$ on detector mismeasurements.
The nominal sample size is reduced by a factor
of three and a half and, again according to our simulation,
the effect of tracking problems is reduced to negligible levels.
We call this the {\em kinematic selected} sample.

	In all three samples the width is extracted with an unbinned maximum
likelihood fit to the energy release distribution and compared with
the simulation's generated value to determine a bias which is
then applied to the data.
These three different approaches yield consistent values for
the width of the $D^{\ast +}$ giving us confidence that our simulation
accurately models our data.

\section{CLEO Detector and Data Samples}

	The CLEO detector has been described in detail elsewhere.  All of
the data used in this analysis are taken with the detector in its
II.V configuration \cite{CLEO}.  This work mainly depends on the 
tracking system
of the detector which consists of a three layer, double sided silicon
strip detector, an intermediate ten layer drift chamber,
and a large 51 layer helium-propane drift chamber.  All three are in an
axial magnetic field of 1.5 Tesla provided by a superconducting solenoid
that contains the tracking region.  The charged tracks are fit using a
Kalman filter technique that takes into account energy loss as the tracks
pass through the material of the beam pipe and detector \cite{kalman}.

	The data were taken in symmetric $e^+e^-$ collisions at a center of
mass energy around 10 GeV with an integrated luminosity of 9.0/fb provided by
the Cornell Electron-positron Storage Ring (CESR).  The nominal sample
follows the selection of
$D^{\ast +} \to \pi_{\rm slow}^+ D^0 \to K^-\pi^+\pi_{\rm slow}^+$ candidates
used in our $D^0-\bar{D^0}$ mixing analysis\cite{Dmix}.

\begin{figure}
     \centerline{\epsfig{file=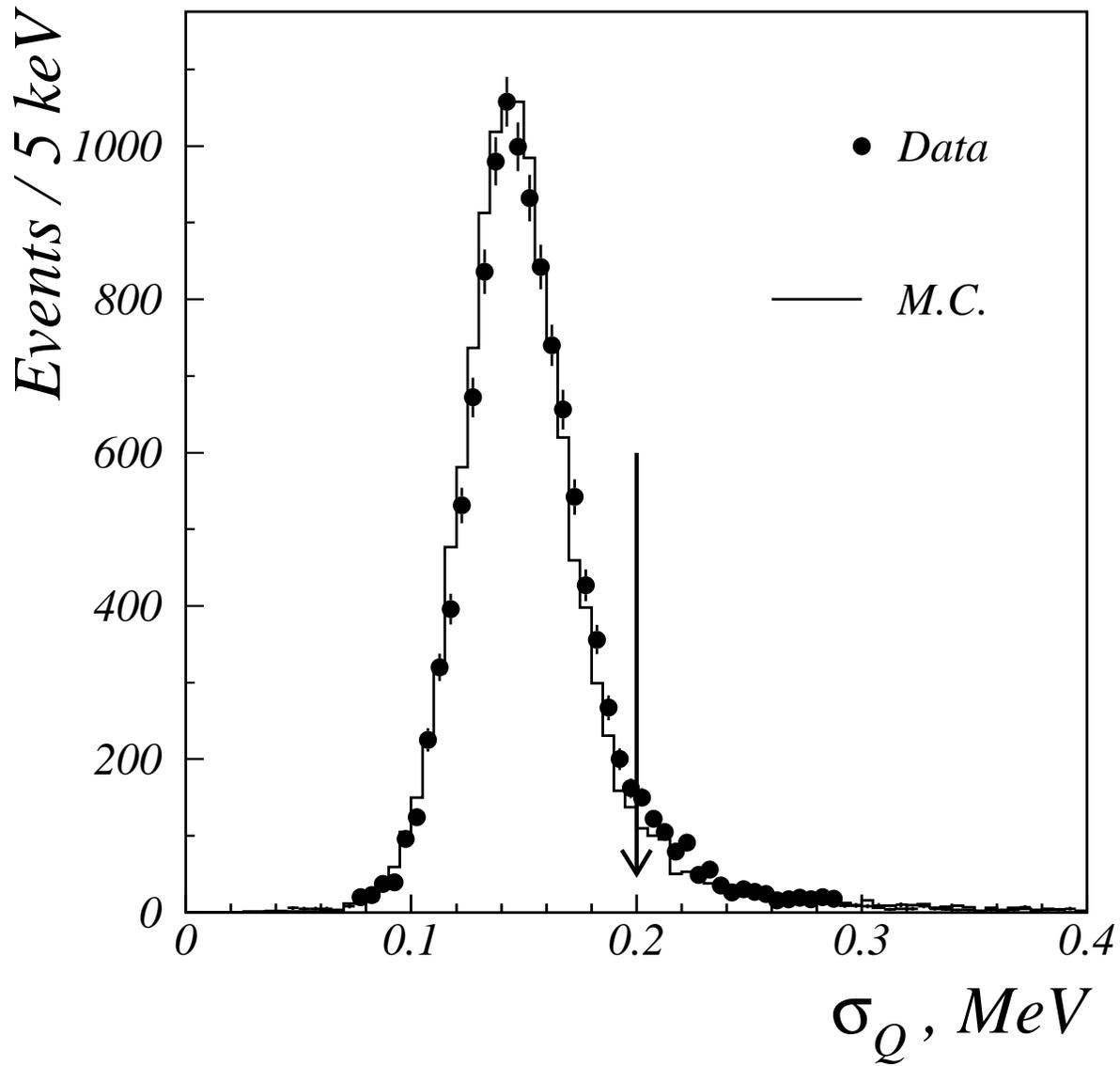,width=\linewidth}}
     \caption{\label{fig:errorcompare} Distribution of $\sigma_Q$,
              the uncertainty on $Q$ as determined from propagating
              track fitting errors.  The arrow indicates a selection
              discussed in the text.}
\end{figure}

	Our reconstruction method takes advantage of the small CESR beam
spot and the kinematics and topology of the
$D^{\ast +} \to \pi^+_{\rm slow} D^0 \to \pi^+_{\rm slow} K^- \pi^+$
decay chain. The $K^-$ and $\pi^+$ are required to form a common vertex.
The resultant $D^0$ candidate momentum vector is then projected back to the
CESR luminous region to determine the $D^0$ production point.  
The CESR luminous region has a Gaussian width $\sim 10\ \mu$m vertically and
$\sim 300\ \mu$m horizontally.  It is well determined
by an independent method\cite{hourglass}.
This procedure determines an accurate
$D^0$ production point for $D^0$'s moving out of the horizontal plane;
$D^0$'s moving within 0.3 radians of the horizontal plane are not
considered.  Then the $\pi_{\rm slow}^+$ track
is refit constraining its trajectory to intersect the $D^0$
production point.  This improves the resolution on the energy release,
$Q = M(K^-\pi^+\pi_{\rm slow}^+) - M(K^-\pi^+) - M(\pi^+)$, by more
than 30\% over simply forming the appropriate invariant masses of the tracks.
The improvement to resolution is essential to our measurement of the width
of the $D^{\ast +}$.  Our resolution is shown in Figure~\ref{fig:errorcompare}
and is typically 150 keV.
The good agreement reflects that the kinematics
and sources of errors on the tracks, such as number of hits
and the effects of multiple scattering in detector material, 
agree between the data and the simulation. 

\begin{figure}
\begin{tabular}{cc}
\epsfxsize=85mm \epsfbox{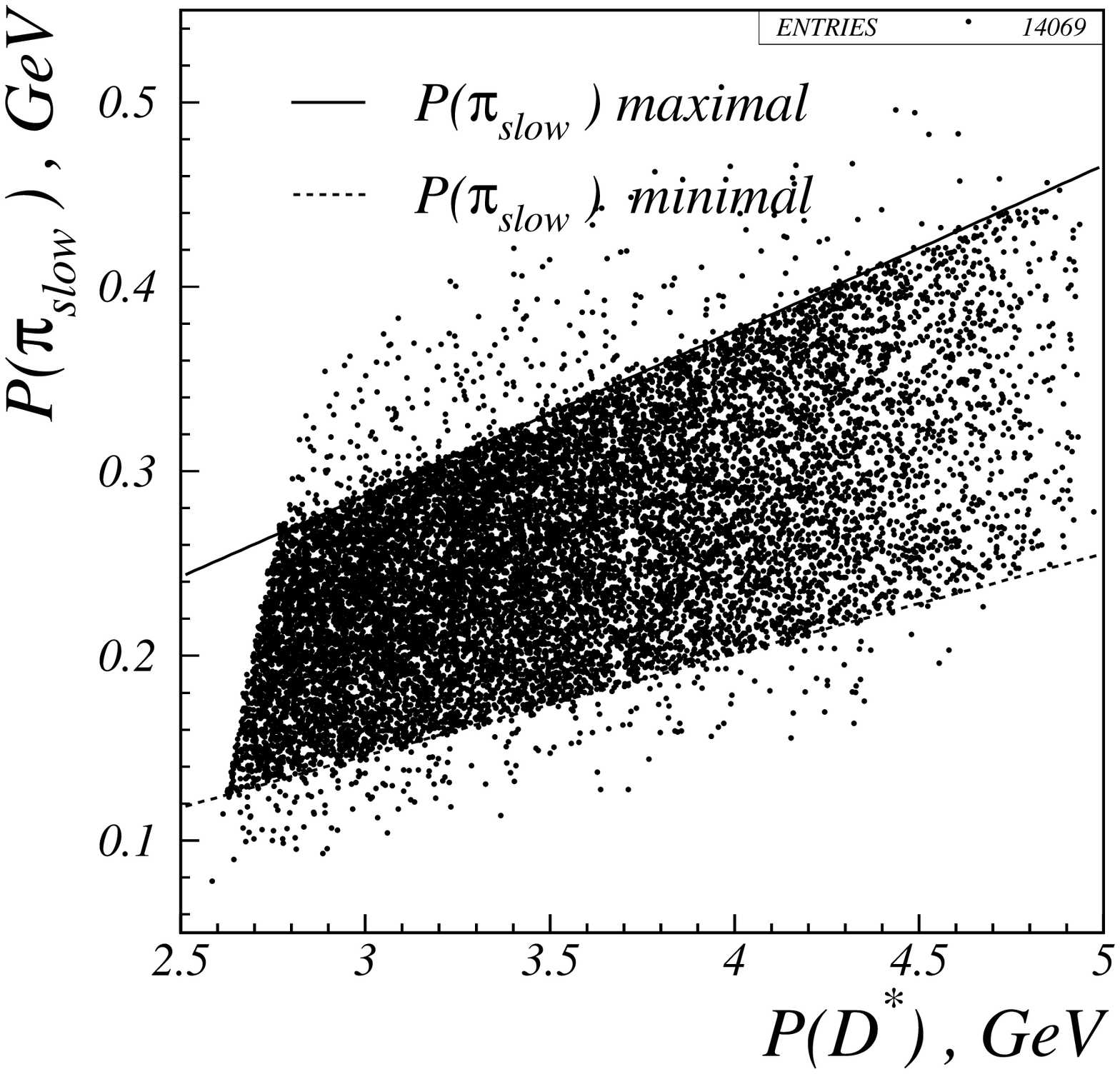} &
\epsfxsize=85mm \epsfbox{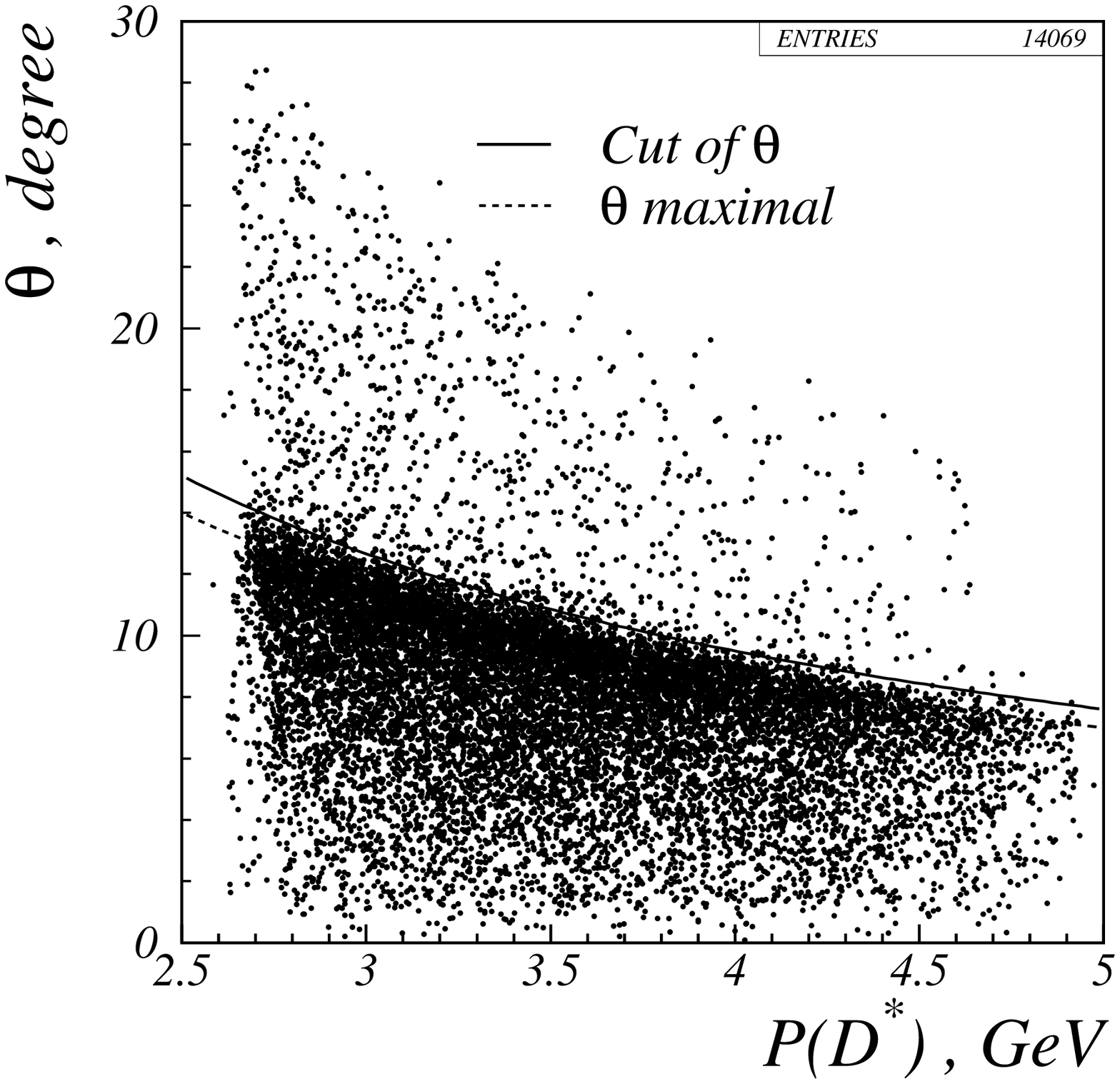}
\end{tabular}
\caption{\label{fig:ppisoft} The $\pi_{\rm slow}^+$ momentum on the left
and the opening angle between $D^0$ and $\pi_{\rm slow}^+$ on the
right both versus the $D^{\ast +}$ momentum in the nominal data sample.}
\end{figure}

	To further improve the quality of reconstruction in our sample, we
apply some kinematic cuts to remove a small amount of misreconstructed
background.
Figure~\ref{fig:ppisoft} shows the distribution of the momentum
of the $\pi_{\rm slow}^+$ as a function of the $D^{\ast +}$ candidate
momentum.  We apply a cut at the kinematic boundary as shown in the
figure.  Figure~\ref{fig:ppisoft} also shows the opening angle $\theta$ between
the $\pi_{\rm slow}^+$ and the $D^0$ candidate as a function of the
$D^{\ast +}$ candidate momentum.  We apply a cut of
$\theta < 38^\circ/P_{D^\ast}[GeV]$ which is just beyond the
kinematic limit to account for resolution smearing.
We also require $\sigma_Q < 200$ keV which removes the long tail in
the error distribution.

	The tracking selected sample makes much more stringent cuts
on the quality of the tracks used to identify the candidates.  All tracks
are required to have hits in both the $r\phi$ and $z$ views
in all three layers of the
silicon strip detector as opposed to the nominal two silicon hits per view.
None of these hits are allowed to be within
2~mm of a silicon wafer edge.  The $D^0$ daughter tracks are required
to have at least 38 of the possible 51 main drift chamber hits
and seven of the ten intermediate drift chamber hits.  The $\chi^2$
per degree of freedom of the fit to these two tracks are limited to
less than 2 in each
of the two drift chambers and 50 in the silicon strip detector.
These selections are designed to remove tracks that have tracking mishaps
or decay in flight.

\begin{figure}
   \begin{minipage}[t]{160mm}
     \epsfxsize=160mm
     \centerline{\epsfbox{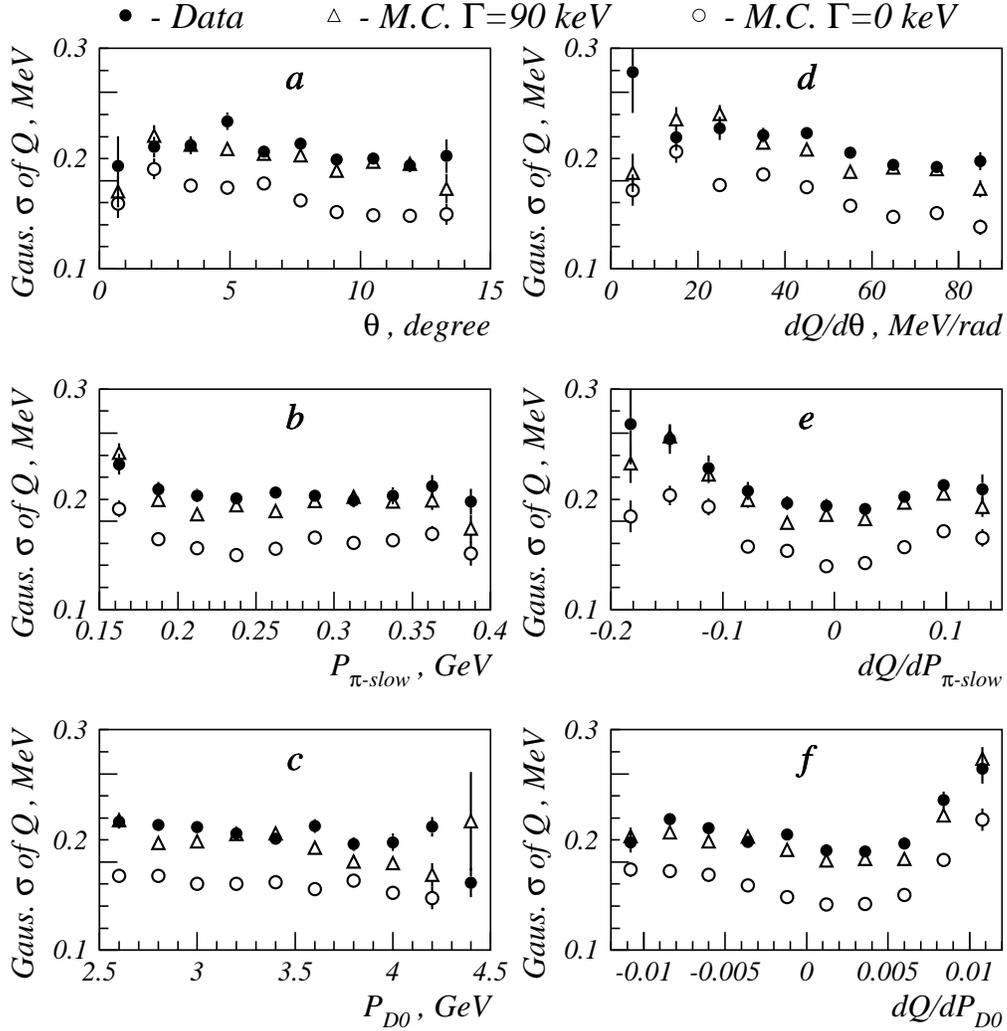}}
     \caption{\label{fig:sigcompare} Gaussian width of $Q$
     distribution versus kinematic parameters and derivatives.
     $\bullet$ -- Data;
     $\circ$--Simulation with $\Gamma_{D^\ast +}=0$;
     $\triangle $--Simulation with $\Gamma_{D^\ast +}=90$ keV. }
   \end{minipage}
\end{figure}

\begin{figure}
   \begin{minipage}[t]{160mm}
     \epsfxsize=160mm
     \centerline{\epsfbox{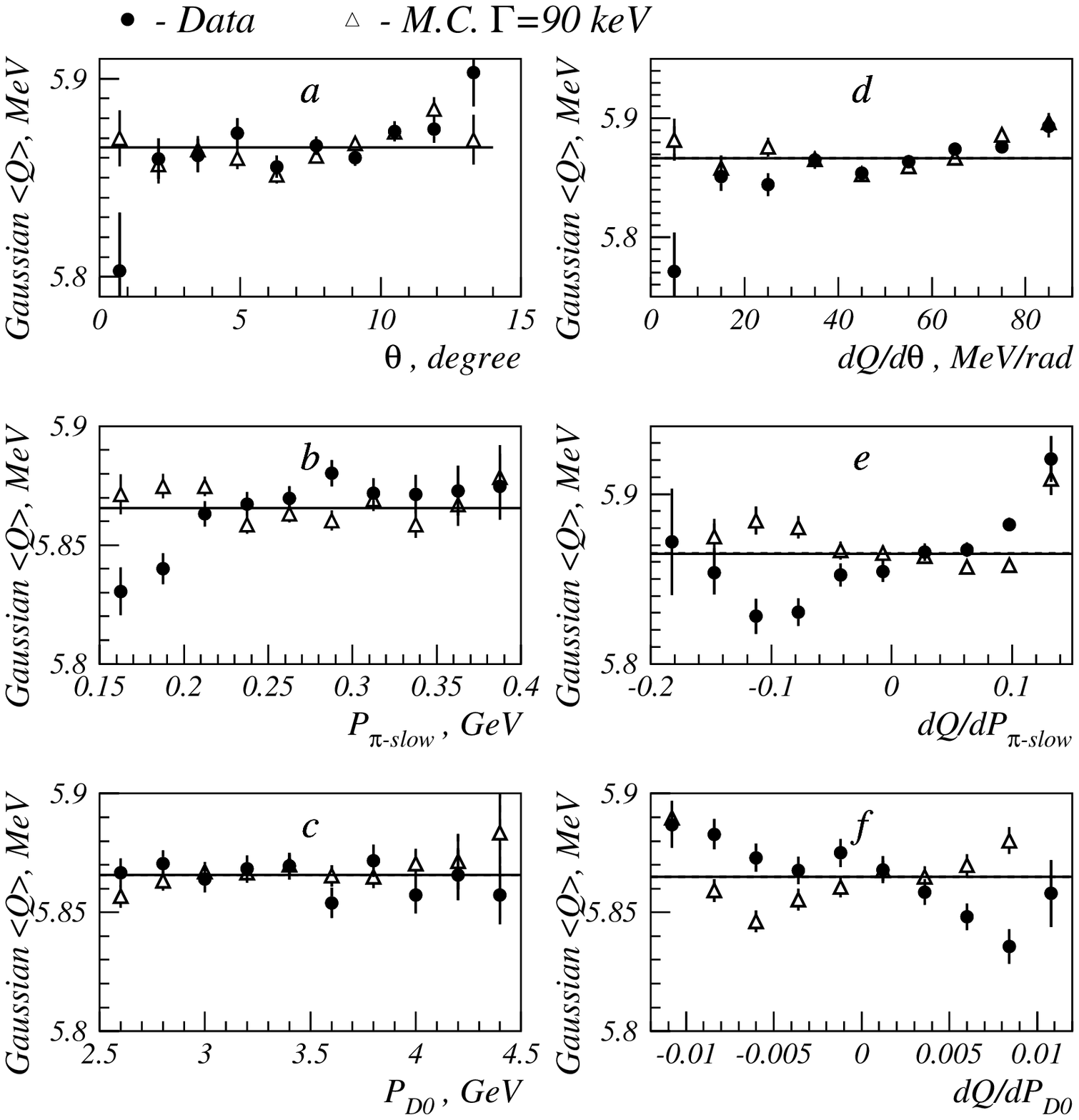}}
     \caption{\label{fig:meancompare} Gaussian mean of $Q$
     distribution versus kinematic parameters and derivatives.
     $\bullet$--Data;
     $\triangle $--Simulation with $\Gamma_{D^\ast +}=90$ keV.
     The horizontal lines show the average value of $Q$ for
     the two samples.  }
   \end{minipage}
\end{figure}

	Since we have no calibration mode we compare the simulation and
the data as a function of kinematic variables of the $D^{\ast +}$ decay.
This will provide another test of the simulation's modeling of the data,
and be the basis of our study of systematic uncertainties in the analysis.
The most important kinematic variables are the ``derivatives'' which
are defined by
\begin{equation}
M^2 = m^2_{D^0}+m^2_{\pi^+_{\rm slow}}+ 2(E_{D^0}E_{\pi^+_{\rm 
slow}}-P_{D^0}P_{\pi^+_{\rm slow}}\cos\theta)\ ,
\end{equation}
\begin{equation}
\beta_{D^0} = P_{D^0}/E_{D^0},
\end{equation}
\begin{equation}
\beta_{\pi^+_{\rm slow}} = P_{\pi^+_{\rm slow}}/E_{\pi^+_{\rm slow}},
\end{equation}
\begin{eqnarray}
\frac{\partial Q}{\partial P_{D^0}} & \equiv &
\frac{E_{\pi^+_{\rm slow}}}{M}(\beta_{D^0} - \beta_{\pi^+_{\rm slow}} 
\cos\theta), \\
\frac{\partial Q}{\partial P_{\pi^+_{\rm slow}}} & \equiv &
\frac{E_{D^0}}{M}(\beta_{\pi^+_{\rm slow}} - \beta_{D^0} \cos\theta), \\
\frac{\partial Q}{\partial \theta} & \equiv &
\frac{P_{D^0} P_{\pi^+_{\rm slow}}}{M}\sin\theta.
\end{eqnarray}
These derivatives test correlations among the basic kinematic
variables, the $D^0$ and $\pi^+_{\rm slow}$ momenta and the opening
angle, $\theta$.  We compare by dividing
the $Q$ distribution into ten slices in each of
the kinematic variables and fitting the ten sub-distributions
of $Q$ to Gaussians.  We display the width and mean
of the ten fits as a function of each of the six kinematic
variables in Figures~\ref{fig:sigcompare} and~\ref{fig:meancompare}.

	The quality of the width comparison (Figure~\ref{fig:sigcompare})
is excellent, with the
simulation generated with an underlying $\Gamma(D^{\ast +}) = 90-100$ keV
accurately predicting the data for all the kinematic variables.  Even when
generated with an underlying $\Gamma(D^{\ast +}) = 0$ keV the simulation
accurately tracks the data's changes as the kinematic variables vary
across their allowed range.

	The quality of the mean comparison (Figure~\ref{fig:meancompare})
is not as good.  The dependence
of the mean of $Q$ is not well modeled versus the $\pi^+_{\rm slow}$
momentum, $\partial Q/ \partial P_{\pi^+_{\rm slow}}$, and 
$\partial Q/\partial P_{D^0}$ by our simulation.  We discuss the consequences
of this imperfect modeling of the data in the section on
systematic uncertainties below.

\begin{figure}
\begin{tabular}{cc}
\epsfxsize=85mm \epsfbox{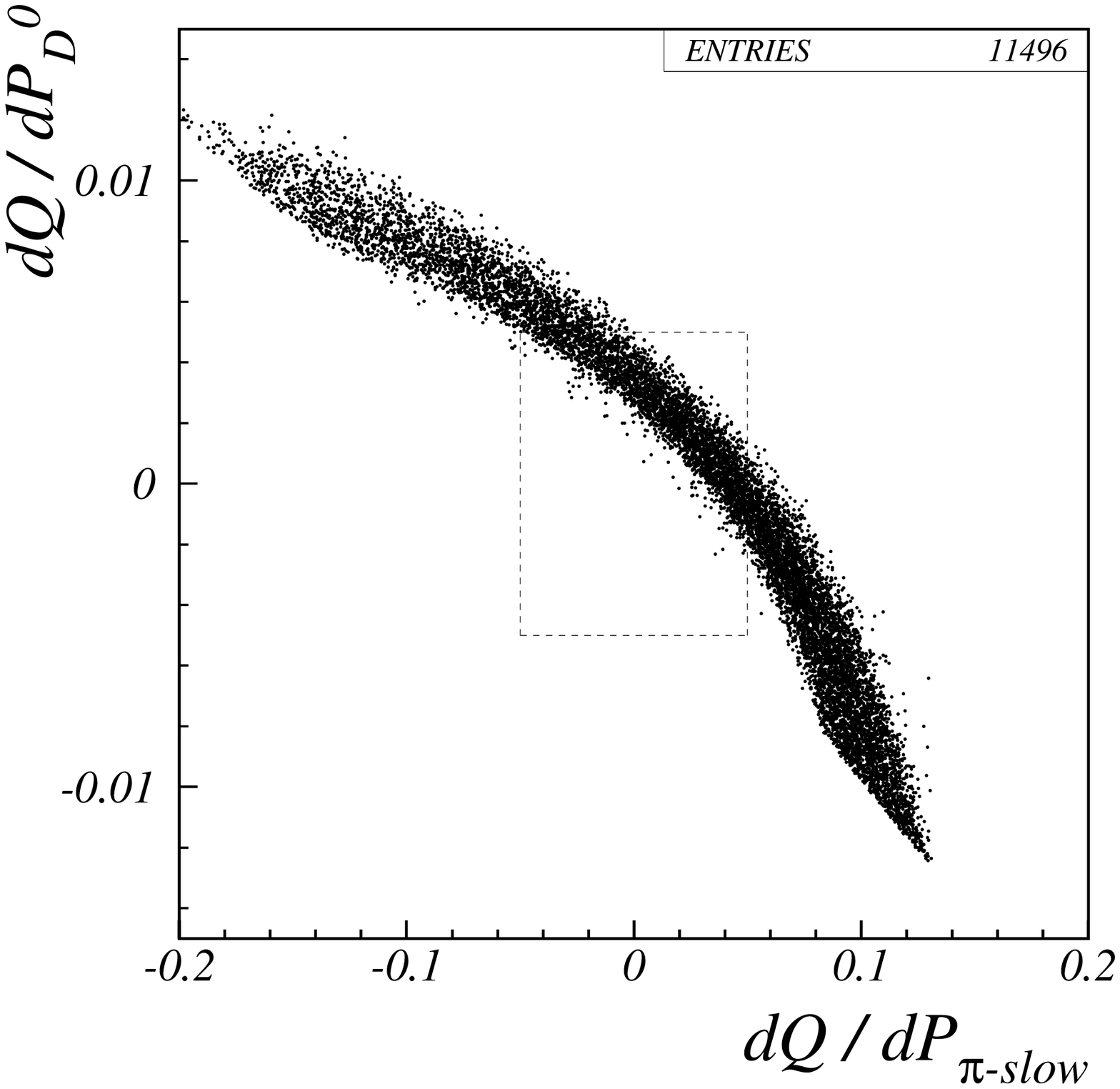} &
\epsfxsize=85mm \epsfbox{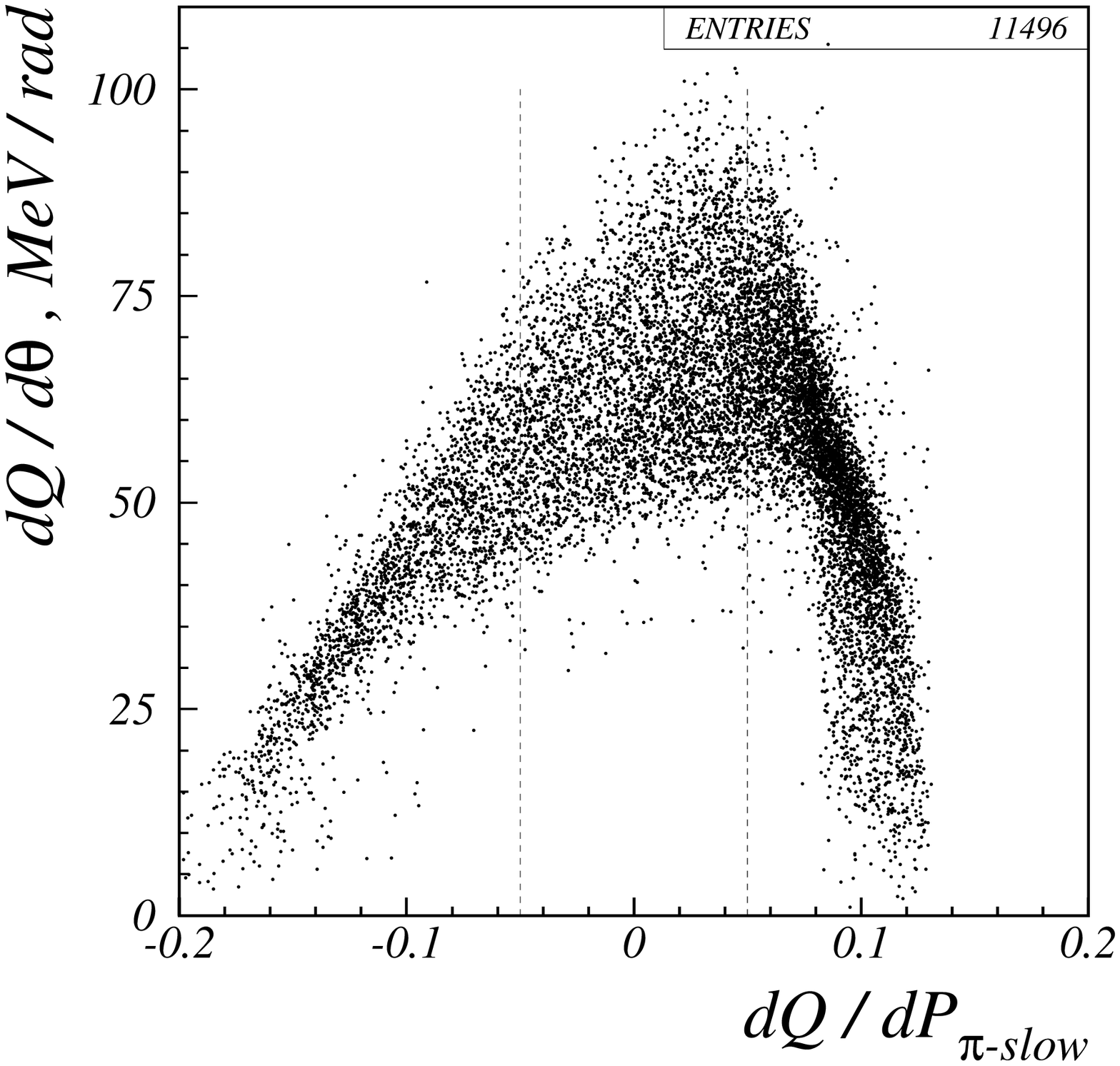} \\
\multicolumn{2}{c} {\epsfxsize=85mm \epsfbox{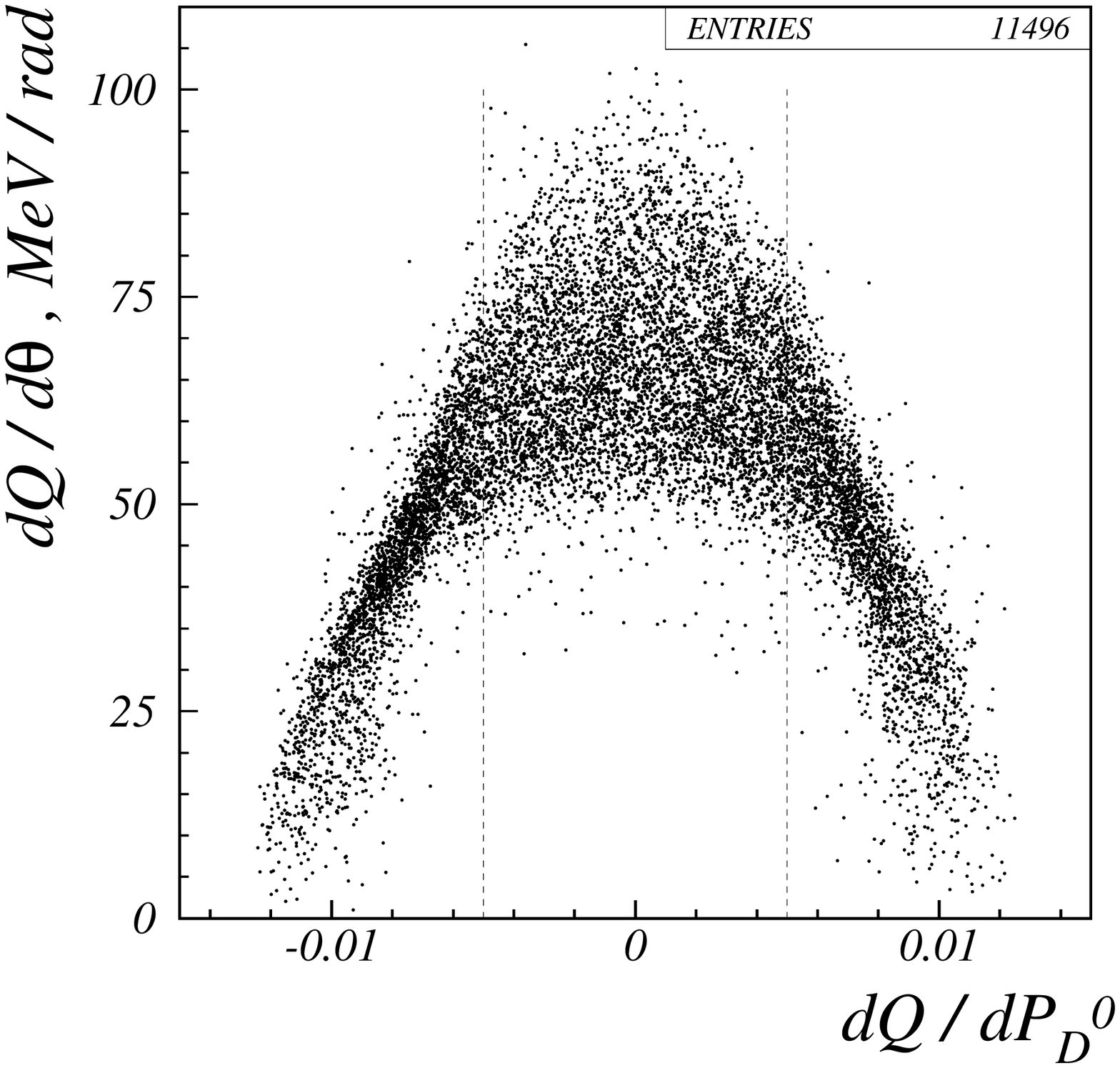}}
\end{tabular}
\caption{\label{fig:derivs} Distributions of $\partial Q /\partial P_{D^0}$
versus $\partial Q / \partial P_{\pi^+_{\rm slow}}$ (top left),
$\partial Q /\partial \theta$
versus $\partial Q / \partial P_{\pi^+_{\rm slow}}$ (top right), and
$\partial Q /\partial \theta$
versus $\partial Q / \partial P_{D^0}$ (bottom) in the data.  The dashed
regions show the selection that defines the kinematic selected sample.}
\end{figure}

	Figure~\ref{fig:derivs} shows
the three derivatives plotted against each other in the data.
Note that if we select
$\partial Q/\partial P_{D^0}$ and $\partial Q/\partial P_{\pi^+_{\rm slow}}$
both to be close to zero we minimize the dependence of $Q$ on the
basic kinematic variables $P_{D^0}$ and $P_{\pi^+_{\rm slow}}$,
and thus minimize the contribution of the kinematic variables to the
width of the $Q$ distribution.  With this selection we are more sensitive
to the underlying width of the $Q$ distribution rather than variations caused
by any mismodeling of $Q$'s dependence on the basic kinematics.
The kinematic selection is defined by
\begin{eqnarray}
\left|\frac{\partial Q}{\partial P_{D^0}}\right| & \leq & 0.005, \\
\left|\frac{\partial Q}{\partial P_{\pi^+_{\rm slow}}}\right| & \leq & 0.05.
\end{eqnarray}

	Table~\ref{tab:data} summarizes the statistics in our three samples.

\section{Fit Description}

	We assume that the intrinsic width of the $D^0$ is negligible,
$\Gamma(D^0) \ll \Gamma(D^{\ast +})$, implying that the width of $Q$
is simply a convolution of the shape given by the $D^{\ast +}$ width and
the tracking system response function.  Thus we
consider the pairs of $Q$ and $\sigma_Q$ for
$D^{\ast +} \to \pi_{\rm slow}^+ D^0 \to K^-\pi^+\pi_{\rm slow}^+$
where $\sigma_Q$ is given for
each candidate by propagating the tracking errors in the kinematic
fit of the charged tracks.  We perform an unbinned maximum likelihood fit
to the $Q$ distribution.

	The shape of the observed $Q$ distribution is assumed to be
given by a signal that is a non-relativistic Breit-Wigner,
\begin{equation}
S = \frac{\Gamma(Q)}{(Q-Q_0)^2+(\Gamma(Q)/2)^2},
\end{equation}
with central value of $Q$, $Q_0$.
We also considered a relativistic Breit-Wigner as a model of the
underlying signal shape, and found negligible changes in the fit
parameters from the non-relativistic shape.
The width of the signal Breit-Wigner is not a constant.  It 
depends on $Q$ and is given by
\begin{equation}
\Gamma(Q) = \Gamma_0  \left(\frac{P}{P_0}\right)^3 \left(\frac{M_0}{M}\right)^2 ,
\label{eq:BW}
\end{equation}
where $\Gamma_0$ is equivalent to $\Gamma(D^{\ast +})$,
$P$ and $M$ are the measured candidate
$\pi_{\rm slow}^+$ or $D^0$ momentum and $D^{\ast +}$ mass in
the $D^{\ast +}$ rest frame and $P_0$ and $M_0$ are
the values computed using $Q_0$.
The effect of the mass term is negligible at our energy. 

	For each candidate the signal shape is convolved with
a resolution Gaussian with width $\sigma_Q$, determined by the tracking
errors, as a model of  our finite resolution.  Figure~\ref{fig:errorcompare}
shows the distribution of $\sigma_Q$ for the data and the simulation.

\begin{figure}
\epsfxsize=\linewidth
\centerline{\epsfbox{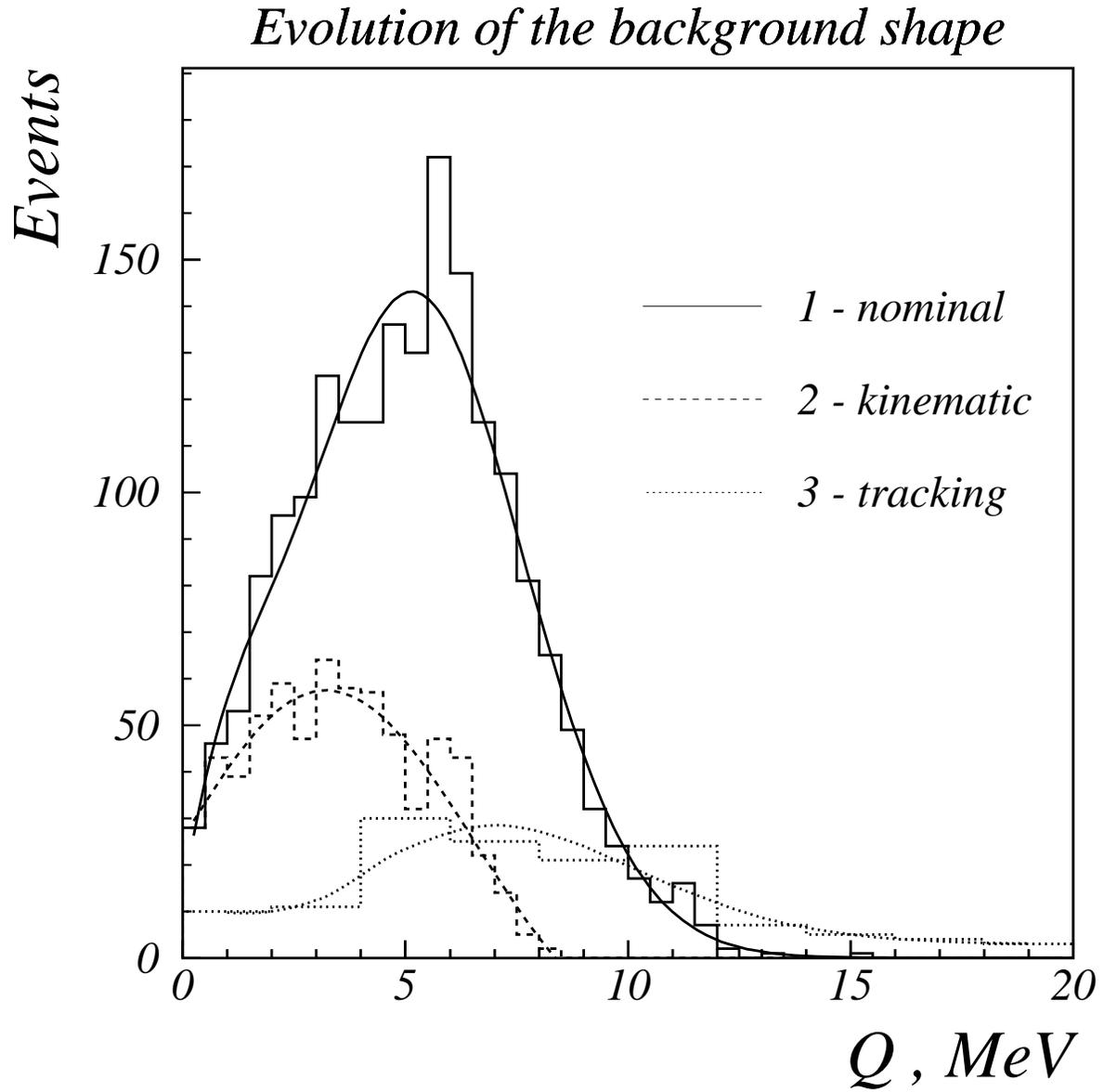}}
\caption{\label{fig:back} Our simulation's prediction of the background
for the three samples discussed in the text.  Also shown are the fits
to third order polynomials that are used in the fits to the data.}
\end{figure}

The fit also includes a background contribution
with a fixed shape.  The shape for the background is taken from fits to the
background prediction of our simulation with a third order polynomial.
The level of the background is allowed to float in our standard
fit.  The predicted background shape and fits are displayed
in Figure~\ref{fig:back}.

\begin{figure}
\epsfxsize=\linewidth
\centerline{\epsfbox{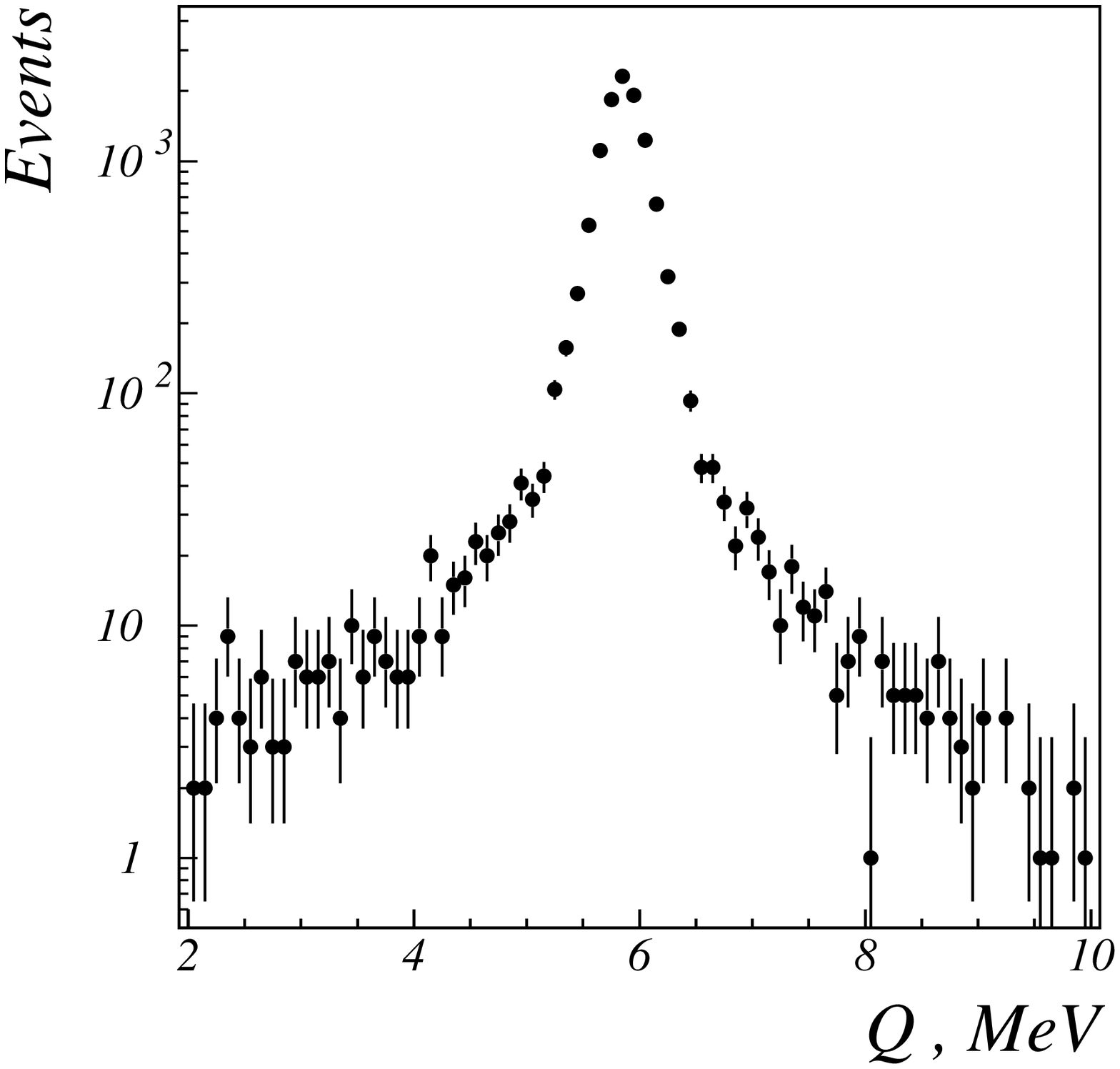}}
\caption{\label{fig:typical} The $Q$ distribution of the nominal
sample in the data.}
\end{figure}

	Figure~\ref{fig:typical} shows the $Q$ distribution for
our nominal data sample.  Note that besides the well measured signal and
the small, slowly varying background, there is also a small
component centered on the signal with a large width.
Therefore we allow a small fraction of the signal $f_{mis}$ to be parametrized
by a single Gaussian resolution function of width $\sigma_{mis}$.
This shape is included in the fit to model the tracking
mishaps which our simulation predicts to be at the 5\% level in the
nominal sample and negligible in both the tracking
and kinematic selected samples.  Typically we constrain the level
of this contribution while allowing $\sigma_{mis}$ to float.

	We have many other parameters of the fit that can be varied
or allowed to float for testing purposes.  We can allow a scale factor
on each candidate's $\sigma_Q$ to model a systematic mistake in our
tracking system caused, for
example, by not properly accounting for the material of the detector.
In our standard fits we
only allow the normalization of the background to float, but we can
either vary the shape as indicated by the simulation or
allow the parameters of the background polynomial to float as a measure
of the small systematic uncertainty due to the background shape.

	Table~\ref{tab:parameters} summarizes the parameters of our
\begin{table}
\caption{Parameters of our fit to the $Q$ distribution}
\begin{center}
\begin{tabular}{|c|c|} \hline
Parameter      & Description \\ \hline
$\Gamma_0$     & Breit Wigner width of $D^{\ast +}$ Signal, 
$\Gamma(D^{\ast +})$ \\
$Q_0$          & Mean of $D^{\ast +}$ Signal \\
$N_s$          & Number of Signal Events \\
$f_{mis}$      & Fraction of Mismeasured Signal\\
$\sigma_{mis}$ & Resolution on Measured $Q$ for Mismeasured Signal\\
$N_b$          & Number of Background Events\\
$k$            & $\sigma_Q$ Scale factor, fixed to 1\\
$B_{1,2,3}$    & Coefficients of Background Polynomial, fixed from 
Simulation \\ \hline
\end{tabular}
\end{center}
\label{tab:parameters}
\end{table}
fit.  Note that the $\sigma_Q$ scale factor $k$ and the background shape
parameters $B_{1,2,3}$ are fixed in our nominal fits.  We minimize the
likelihood function
\begin{equation}
L = 2(N_s + N_b) - 2 \sum _{i=1}^N \log 
[N_s S(Q_i, \sigma_{Qi}; \Gamma_0, Q_0, f_{mis}, \sigma_{mis} ) + 
N_b B(Q_i; B_{1,2,3})],
\end{equation}
where $S$ and $B$ are respectively the signal and background shapes
discussed above.

	The fitter has been extensively tested both numerically
and with input from our full simulation.  We find that the fitter
performs reliably giving normal distributions for the floating
parameters and their uncertainties.  It also reproduces the input
$\Gamma(D^{\ast +})$ from 0 to 130 keV.  Its behavior on
each of the three data samples: nominal; tracking selected; and
kinematic selected in the full simulation is discussed below.
We note that if all the parameters are allowed to vary simultaneously
there is strong correlation among the intrinsic width $\Gamma_0$,
the fraction of mismeasured events $f_{mis}$, and the $\sigma_Q$ scale
factor $k$, as one would expect.
Thus our nominal fit holds $k$ fixed, but in our systematic
studies we either fix one of the three or provide a constraint with
a contribution to the likelihood if the parameter varies from its
nominal value.

\section{Fit Results}

	As a preliminary test to fitting the data we run the complete analysis
on a fully simulated sample that has about ten times the data
statistics and is generated with a range of underlying $\Gamma(D^{\ast +})$
from 0~to 130~keV.  We do this for nominal, tracking, and kinematic
selected samples.  For the nominal sample we note that the fit is not
stable if all the parameters are left to vary freely.  We have
found that if we constrain the fraction of mismeasured signal
to $5.3 \pm 0.5$\% as indicated by the simulation over the range
of generated widths of the $D^{\ast +}$ then we get a stable result.
This constraint makes the
fit to the simulated nominal sample have no significant offset between
the generated and measured values for the width of the $D^{\ast +}$.
The tracking and kinematic selected samples have a negligible amount
of mismeasured signal according to the simulation and in fits
to these samples we fix $f_{mis}$ to zero.
These simulated samples are also consistent with no offset between
the generated and measured values for the width of the $D^{\ast +}$.
We also note that in all three simulated samples there are no trends
in the difference between measured and generated width as a function
of the generated width; the offset is consistent with zero as a function of
the generated width of the $D^{\ast +}$.
Table~\ref{tab:data} summarizes this simulation
study.  We will apply these offsets to the fit value that we obtain
from the data.

\begin{figure}
\epsfxsize=\linewidth
   \centerline{\epsfbox{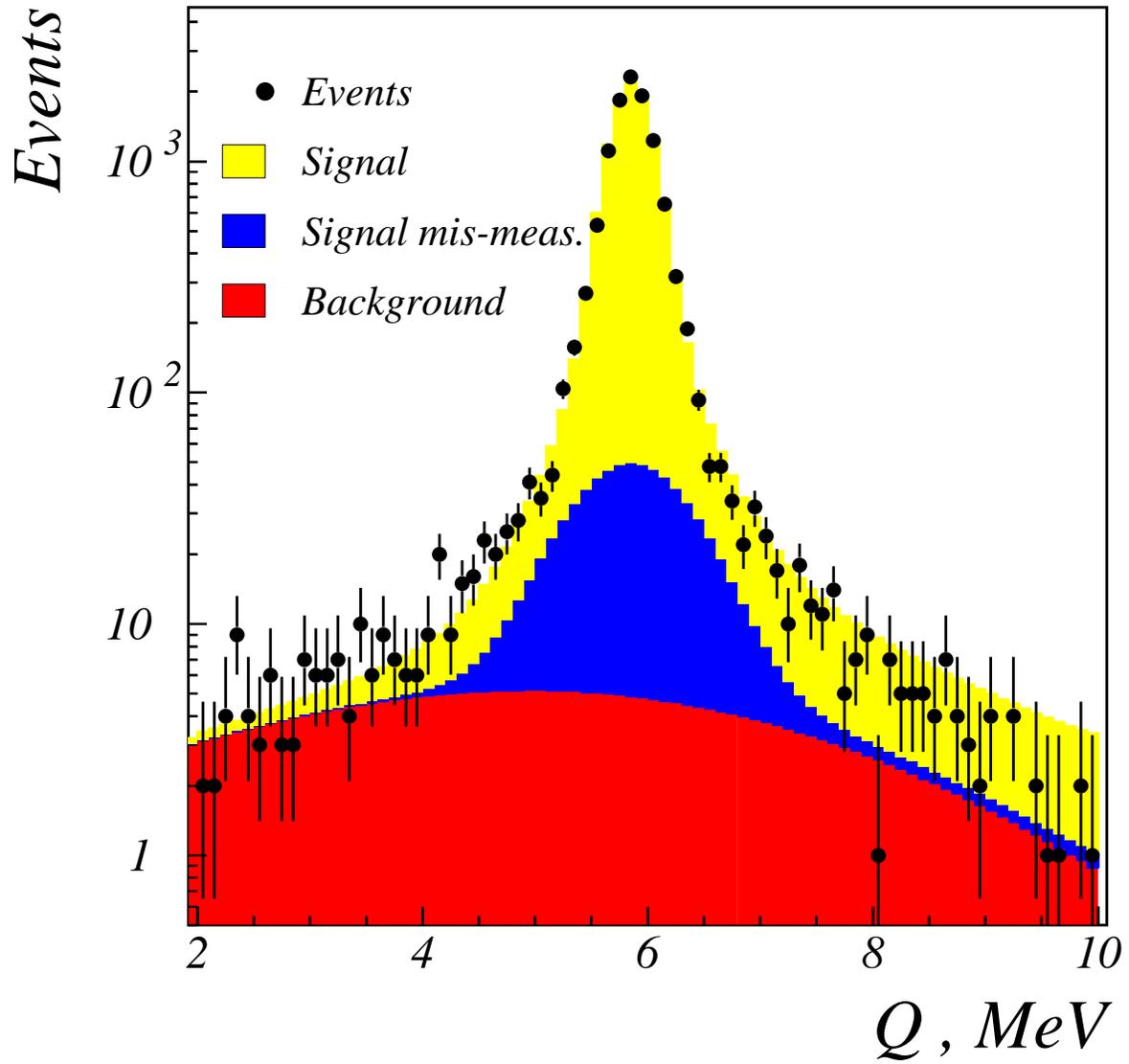}}
   \caption{\label{fig:nomfit} Fit to nominal data sample.  The different
contributions to the fit are shown by different colors.}
\end{figure}

\begin{figure}
\epsfxsize=\linewidth
   \centerline{\epsfbox{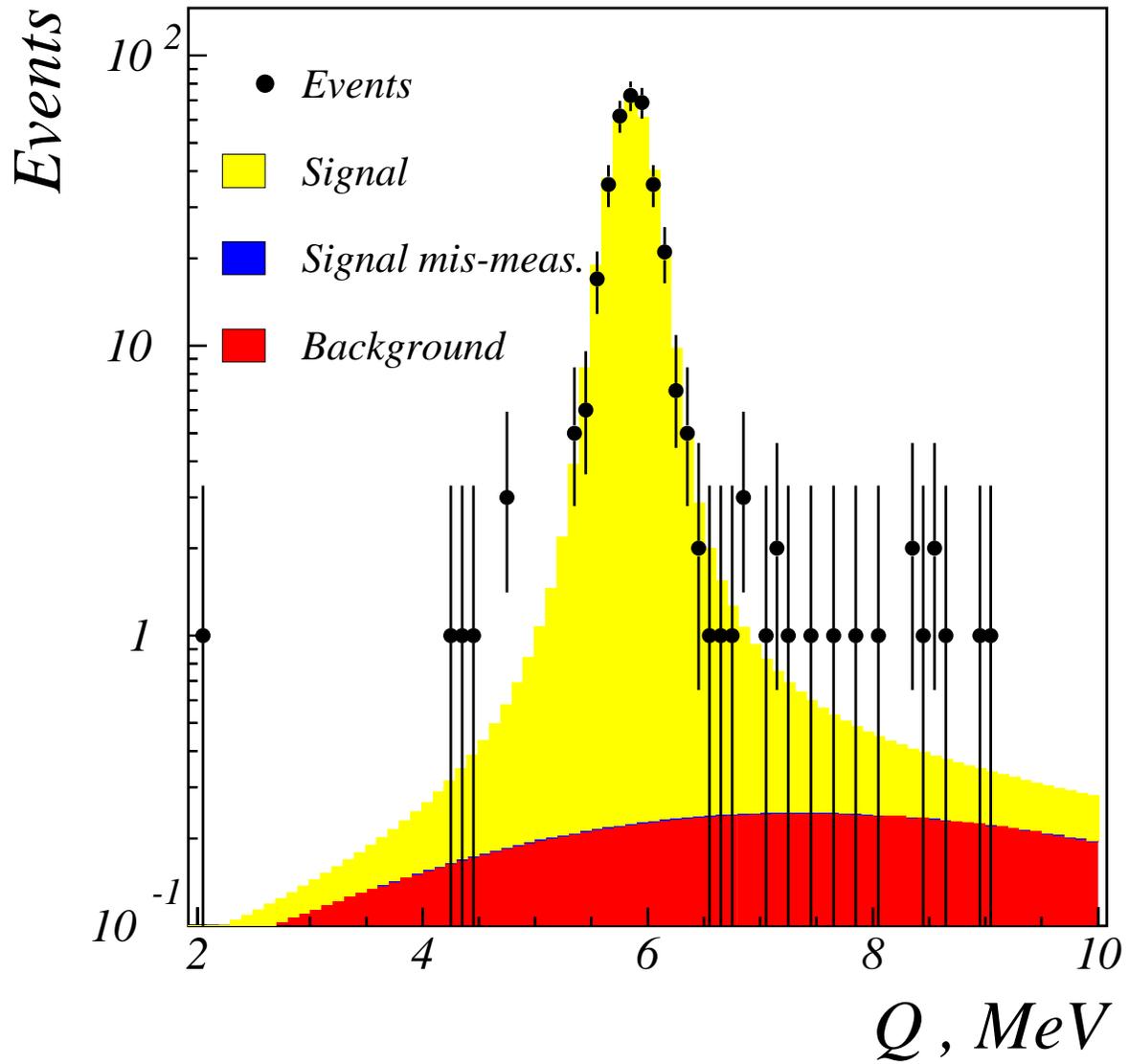}}
   \caption{\label{fig:platinumfit} Fit to tracking selected data sample.
   The different
   contributions to the fit are shown by different colors.}
\end{figure}

\begin{figure}
\epsfxsize=\linewidth
   \centerline{\epsfbox{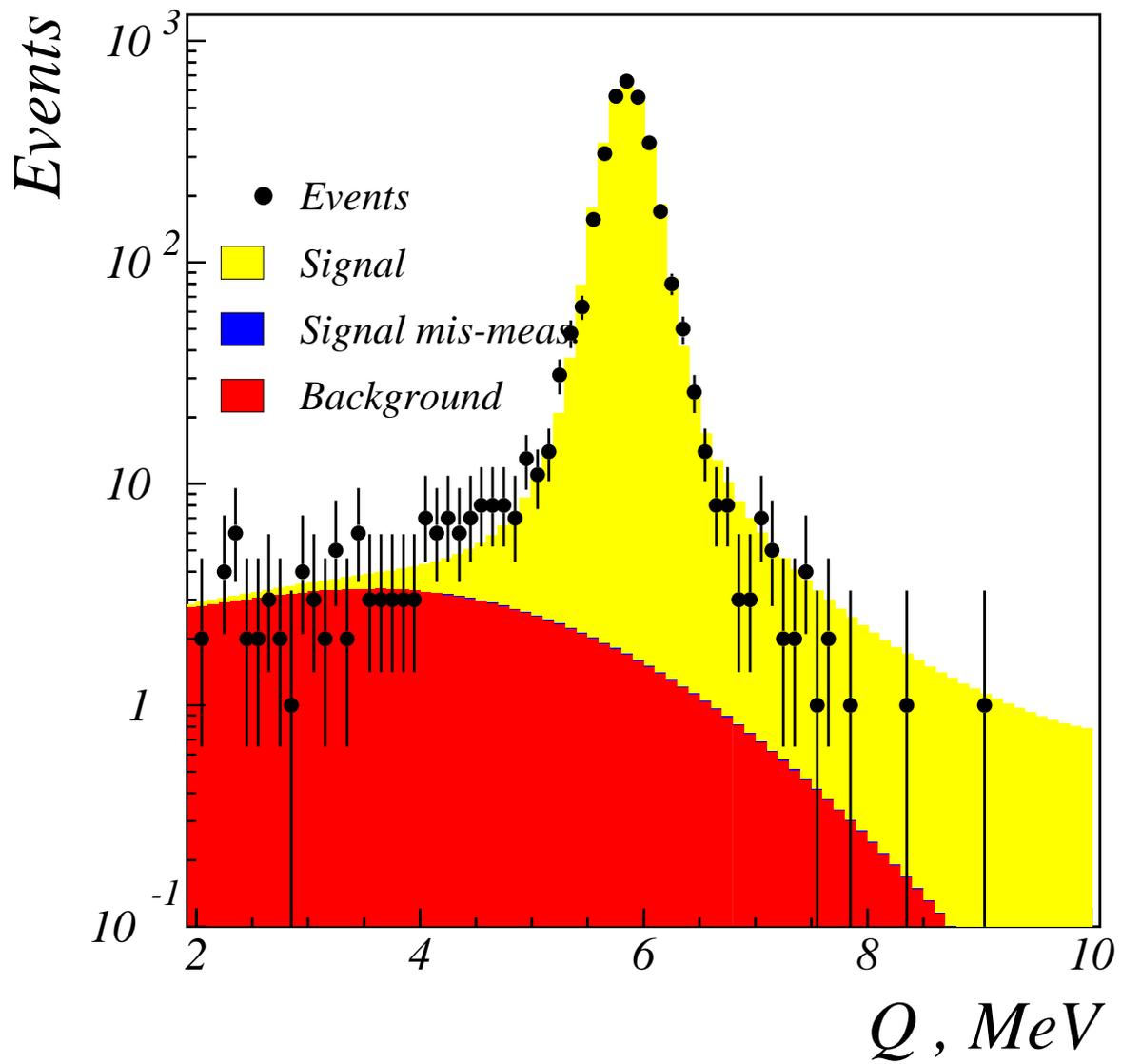}}
   \caption{\label{fig:goldfit} Fit to kinematic selected data sample.
   The different
   contributions to the fit are shown by different colors.}
\end{figure}

\begin{figure}
\begin{tabular}{ccc}
\epsfxsize=55mm
\epsfbox{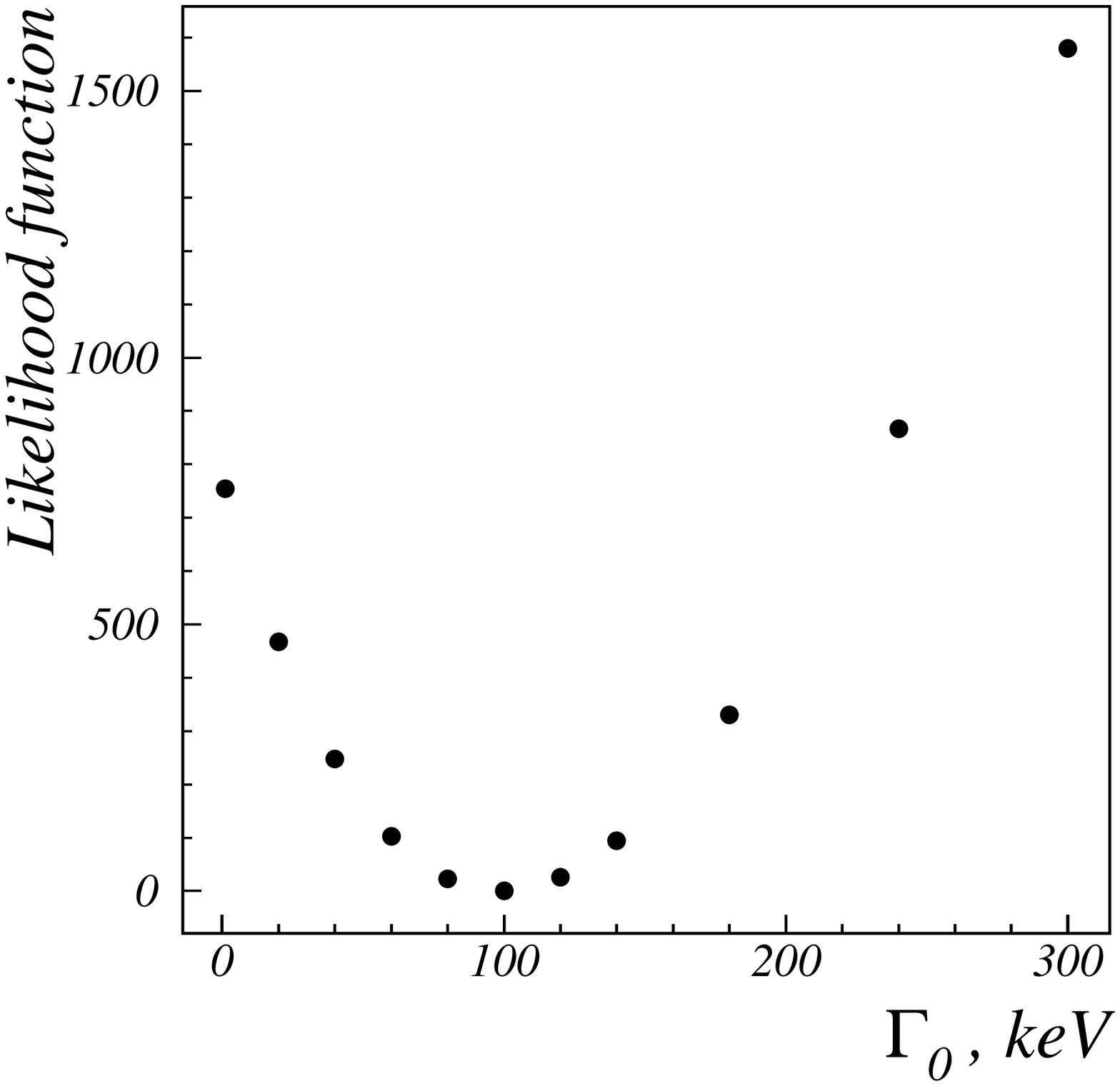} &
\epsfxsize=55mm
\epsfbox{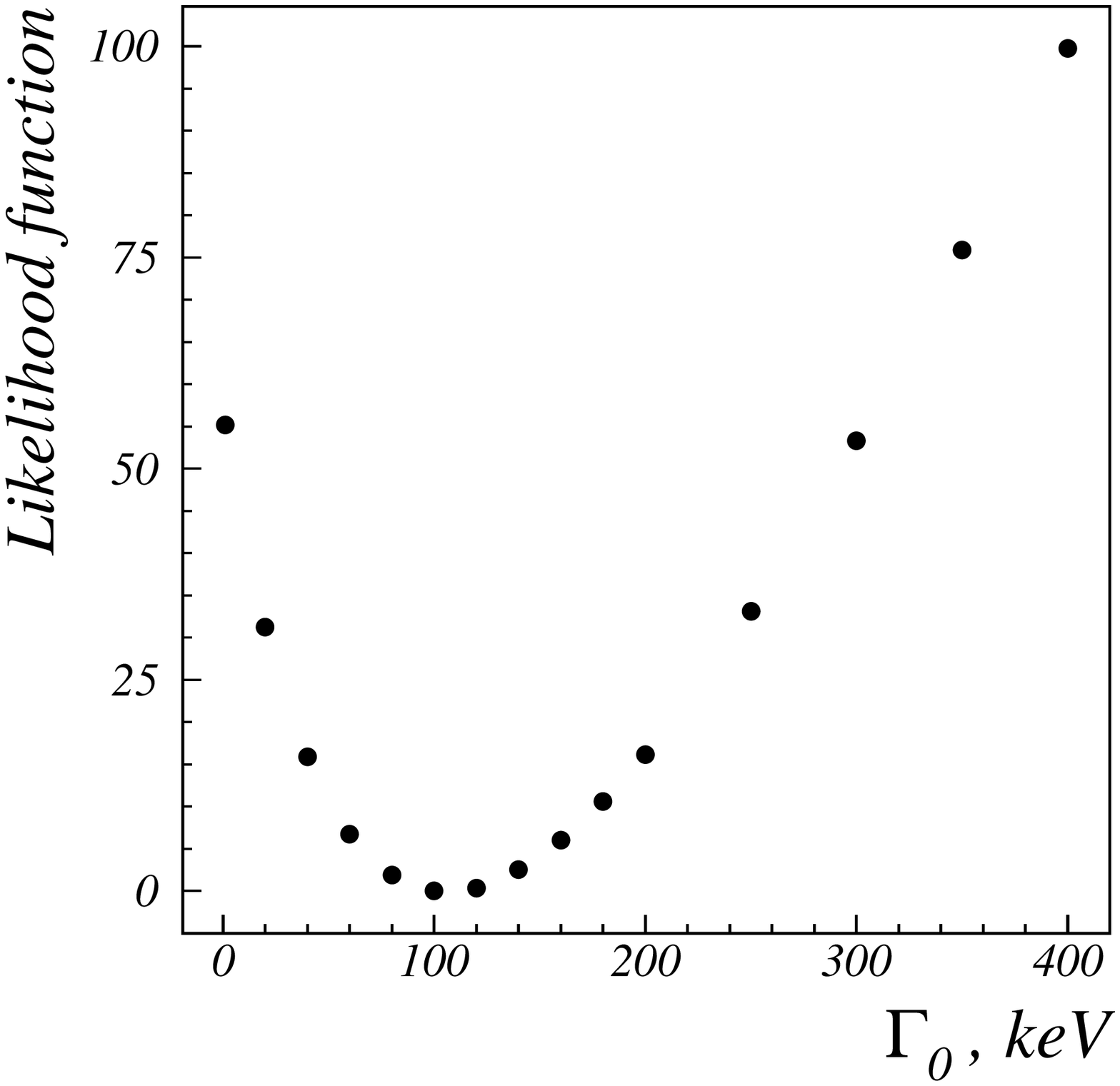} &
\epsfxsize=55mm
\epsfbox{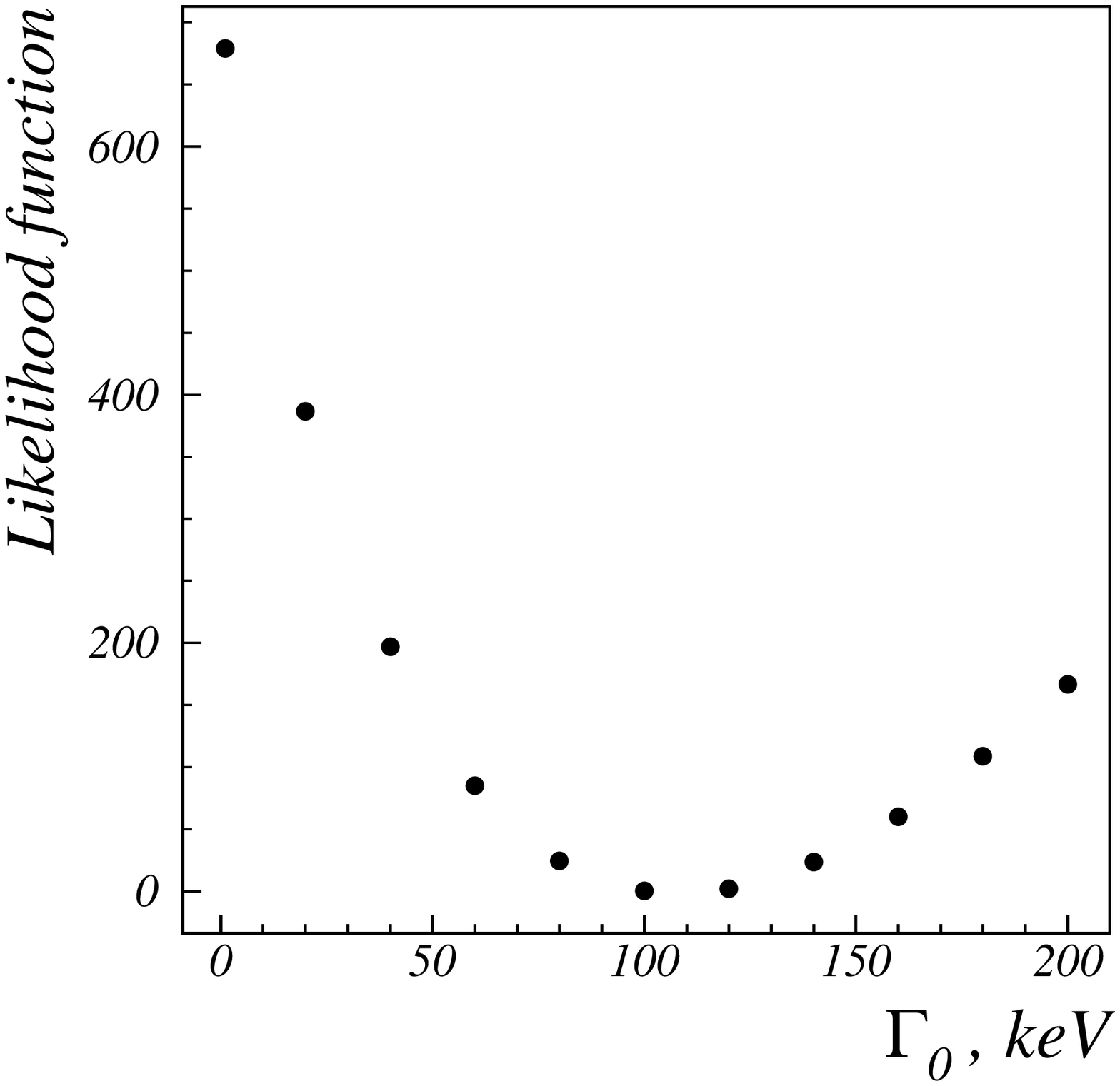} 
\end{tabular}
\caption{\label{fig:like}
Likelihood function versus measured $D^{\ast +}$ width for the nominal (left),
tracking (center), and kinematic (right) selected data samples.}
\end{figure}

	Figures~\ref{fig:nomfit}, \ref{fig:platinumfit}, and \ref{fig:goldfit}
respectively display the fit to the nominal, tracking, and kinematic selected
data samples.  The results of the fits are summarized in Table~\ref{tab:fit}.
\begin{table}
\caption{Results of the fits described in the text.  The fit parameters
are summarized in Table~\ref{tab:parameters}.  The uncertainties are
statistical.}
\begin{center}
\begin{tabular}{|c|c|c|c|} \hline
                      & \multicolumn{3}{|c|}{Sample} \\  \cline{2-4}
Parameter            & Nominal          & Tracking          & 
Kinematic \\ \hline
$\Gamma_0$ (keV)     & $98.9 \pm 4.0$   & $106.0 \pm 19.6$  & $108.1 
\pm  5.9$ \\
$Q_0$ (keV)          & $5853 \pm 2$     & $5854 \pm 10$     & $ 5850 \pm 4$ \\
$N_s$                & $11207 \pm 109$  & $353 \pm 20$      & $3151 \pm 57$ \\
$f_{mis}$ (\%)       & $5.3 \pm 0.5$    & NA                & NA \\
$\sigma_{mis}$ (keV) & $508 \pm 39 $    & NA                & NA \\
$N_b$                & $289 \pm 31 $    & $15 \pm 7$        & $133 \pm 16$ \\ \hline
\end{tabular}
\end{center}
\label{tab:fit}
\end{table}
Correlations among the floating parameters of the fit are negligible.
Figure~\ref{fig:like}
displays the likelihood as a function of the width of the $D^{\ast +}$
for the fits to the three data samples.

	The agreement is excellent among the three fits, and when the offsets
from Table~\ref{tab:data} are applied we obtain
\begin{eqnarray}
{\rm Nominal\ Sample}     & \Gamma(D^{\ast +}) = 96.2 \pm 4.0\ {\rm keV}, \\
{\rm Tracking\ Selected}  & \Gamma(D^{\ast +}) = 104  \pm 20\ {\rm 
keV}, {\rm and}\\
{\rm Kinematic\ Selected} & \Gamma(D^{\ast +}) = 103.8 \pm 5.9\ {\rm keV}.
\end{eqnarray}
The data sample and results are summarized in Table~\ref{tab:data}.
\begin{table}
\caption{Summary of our data sample, simulation biases, and fit results.}
\begin{center}
\begin{tabular}{|c|c|c|c|} \hline
                     & \multicolumn{3}{|c|}{Sample} \\ \cline{2-4}
Parameter            & Nominal          & Tracking          & Kinematic \\ \hline
Candidates           & 11496            & 368               & 3284 \\
Background Fraction (\%)
                     & $2.51 \pm 0.27$  & $4.1 \pm 1.9$     & $4.05 \pm 0.49$ \\
$\Gamma_{\rm fit} - \Gamma_{\rm generated}$ (keV)
                     & $2.7 \pm 2.1$    & $1.7 \pm 6.4$     & $4.3 \pm 3.1$ \\
Fit $\Gamma_0$ (keV) &  $98.9 \pm 4.0$  & $106.0 \pm 19.6$ & $108.1 \pm  5.9$ \\       
$D^{\ast +}$ Width (keV) & $96.2 \pm 4.0$
                         & $104 \pm 20$
                         & $103.8 \pm 5.9$ \\ \hline
\end{tabular}
\end{center}
\label{tab:data}
\end{table}
The uncertainties are only statistical.  We discuss systematic
uncertainties in the next section.

\section{Systematic Uncertainties}

	We discuss the sources of systematic uncertainties on our measurements
of the width of the $D^{\ast +}$ in
the order of their size.  The most important contribution is the variation
of the average $Q$ as a function of the kinematic parameters of
the $D^{\ast +}$ decay as shown in Figure~\ref{fig:meancompare}.  We
estimate this uncertainty by repeating the fits described above  in three
bins for each of the six kinematic parameters and taking the uncertainty as
the largest observed variation from the nominal values in Table~\ref{tab:fit}.
We obtain uncertainties of $\pm 16$ and $\pm 15$ keV on $\Gamma(D^{\ast +})$
and $Q_0$ respectively.

	The next most important contribution
comes from any mismodeling of $\sigma_Q$'s
dependence on the kinematic parameters.  We estimate this by varying
our cut on $\sigma_Q$ from 75 to 400 from the nominal 200 keV and repeating
our analysis with all parameters fixed except allowing the error scale
factor $k$ to vary freely.   This indicates that the resolution is correct
to $\pm 4$\%, and we then repeat our standard analysis with $k$ fixed
at 0.96 and 1.04.  We find uncertainties of $\pm 11$, $\pm 9$,
and $\pm 7$ keV on
$\Gamma(D^{\ast +})$ for the nominal, tracking, and kinematic selected
sample.  For $Q_0$ this uncertainty is negligible except in the tracking
selected sample where it is $\pm 4$ keV.

	We take into account correlations among the
less well measured parameters of the fit, such as $k$, $f_{mis}$, and
$\sigma_{mis}$,
by fixing each parameter at $\pm 1\sigma$ from their central fit values,
repeating the fit, and adding in quadrature the variation in
the width of the $D^{\ast +}$ and $Q_0$ from their central values.
We find uncertainties of $\pm 8$, $\pm 9$, and $\pm 9$ keV on
the width of the $D^{\ast +}$ for the nominal, tracking, and kinematic selected
sample, and respectively $\pm 3$, $\pm 4$, and $\pm 5$ keV on $Q_0$.

	We have studied in the simulation the sources of
mismeasurement that give rise
to the resolution on the width of the $D^{\ast +}$ by replacing the
measured values with the generated values for various kinematic
parameters of the decay products.  We have then compared
these uncertainties with analytic expressions for the uncertainties.
The only source of resolution that we cannot account for in
this way is a small distortion of the kinematics of the event
caused by the algorithm used to reconstruct the $D^0$ origin point
described above.  This contributes an uncertainty $\pm 4$ keV on
the width of the $D^{\ast +}$ and $\pm 2$ keV on $Q_0$.

	We consider uncertainties from the background shape by allowing the
coefficients of the background polynomial to float. 
We observe changes on the width of $\pm 4$ keV for the nominal sample
and $\pm 2$ keV for the tracking and kinematic selected samples.
We have also released our kinematic selection cuts which causes the
background to increase by a large factor.  This causes a change which
is small compared to allowing the coefficients of the background shape
polynomial to float.  Variations in the background have a negligible
effect on $Q_0$.
   
	Minor sources of uncertainty are from the width offsets derived
from our simulation and given in Table~\ref{tab:data}, and
our digitized data storage format which saves track parameters
with a resolution of 1~keV and contributes an uncertainty
of $\pm 1$ keV on the width of the $D^{\ast +}$ and $Q_0$.

	An extra and dominant source of uncertainty on $Q_0$ is the energy
scale of our measurements.  We are still evaluating the size of this
contribution.

	Table~\ref{tab:systematic} summarizes the systematic
\begin{table}
\caption{Systematic uncertainties on the width of the $D^{\ast +}$ and $Q_0$}
\begin{center}
\begin{tabular}{|c|c|c|c|c|c|c|} \hline
                         & \multicolumn{6}{|c|}{Uncertainties in keV} \\
                         & \multicolumn{6}{|c|}{Sample} \\ \cline{2-7}
                         & \multicolumn{2}{|c|}{Nominal}
                         & \multicolumn{2}{|c|}{Tracking}
                         & \multicolumn{2}{|c|}{Kinematic} \\ \hline

Source                    & $\delta \Gamma(D^{\ast +})$ & $\delta Q_0$
                           & $\delta \Gamma(D^{\ast +})$ & $\delta Q_0$
                           & $\delta \Gamma(D^{\ast +})$ & $\delta Q_0$ \\ \hline
Running of $Q$            & 16 & 15   & 16 & 15   & 16 & 15 \\
Mismodeling of $\sigma_Q$ & 11 & $<1$ &  9 &  4   &  7 & $<1$ \\
Fit Correlations          &  8 &  3   &  9 &  4   &  9 &  5 \\
Vertex Reconstruction     &  4 &  2   &  4 &  2   &  4 &  2 \\
Background Shape          &  4 & $<1$ &  2 & $<1$ &  2 & $<1$ \\
Offset Correction         &  2 & NA   &  6 & NA   &  3 & NA \\
Data Digitization         &  1 &  1   &  1 &  1   &  1 & 1  \\ \hline
Quadratic Sum             & 22 & 15   & 22 & 16   & 20 & 16 \\ \hline
\end{tabular}
\end{center}
\label{tab:systematic}
\end{table}
uncertainties on the width of the $D^{\ast +}$ and $Q_0$.

\section{Conclusion}

	We have measured the width of the $D^{\ast +}$ by studying the
distribution of the energy release in $D^{\ast +} \to D^0 \pi^+$ followed
by $D^0 \to K^- \pi^+$ decay.  We have done this in three separate samples,
one that is minimally selected, a second that reduces poorly
measured tracks
due to misassociated hits and non-Gaussian scatters in the detector
material, and a third that takes advantage of the kinematics of the decay
chain to reduce dependence on mismeasurements of kinematic parameters.
The resolution on the energy release is well modeled by our simulation,
with agreement between the sources of the resolution as predicted
by the simulation and analytic calculations.  
The largest sources of uncertainty are
imperfect modeling of the dependence of the mean energy release on the
kinematics of the decay chain, 
the simulation of the error on the energy release,
and correlations among the parameters of the fit to the energy release
distribution.  
With our estimate of the systematic uncertainties for each of the
three samples being essentially the same we chose to report the result
for the sample with the smallest statistical uncertainty, the minimally
selected sample, and obtain
\begin{equation}
\Gamma(D^{\ast +}) = 96 \pm 4 \pm 22\ {\rm keV},
\label{eq:result}
\end{equation}
where the first uncertainty is statistical and the second is systematic.
We note that if we form an average value taking into account the
statistical correlations among our three measures we get a result
that is nearly identical with Equation~\ref{eq:result} since
the average is dominated by the result with the smallest statistical
uncertainty.  This result is preliminary.

This preliminary measurement is the first of the width of
the $D^{\ast +}$, and our
measurement corresponds to a strong coupling\cite{pred}
\begin{equation}
g = 17.9 \pm 0.3 \pm 1.9.
\end{equation}
This is consistent with theoretical predictions based on HQET and
relativistic quark models, but higher than predictions based on QCD
sum rules.  

\section*{Acknowledgements}

We gratefully acknowledge the effort of the CESR staff in providing us with
excellent luminosity and running conditions.
M. Selen thanks the PFF program of the NSF and the Research Corporation, 
and A.H. Mahmood thanks the Texas Advanced Research Program.
This work was supported by the National Science Foundation, the
U.S. Department of Energy, and the Natural Sciences and Engineering Research 
Council of Canada.

\end{document}